\documentclass[10pt]{revtex4}
\usepackage{graphicx}
\usepackage{amsmath,amssymb}
\usepackage{mathtools}

\begin{document}

\title{Jordan in The Church of The Higher Hilbert Space: \\Entanglement and Thermal Fluctuations}
\author{Vlatko Vedral}
\affiliation{Clarendon Laboratory, University of Oxford, Parks Road, Oxford OX1 3PU, United Kingdom\\Centre for Quantum Technologies, National University of Singapore, 3 Science Drive 2, Singapore 117543\\
Department of Physics, National University of Singapore, 2 Science Drive 3, Singapore 117542}

\begin{abstract}
I revisit Jordan's derivation of Einstein's formula for energy fluctuations in the black body in thermal equilibrium. This formula is usually taken to represent the unification of the wave and the particle aspects of the electromagnetic field since the fluctuations can be shown to be the sum of wave-like and particle-like contributions. However, in Jordan's treatment there is no mention of the Planck distribution and all averages are performed with respect to pure number states of radiation (mixed states had not yet been discovered!).  The chief reason why Jordan does reproduce Einstein's result despite not using thermal states of radiation is that he focuses on fluctuations in a small (compared to the whole) volume of the black body. The state of radiation in a small volume is highly entangled to the rest of the black body which leads to the correct fluctuations even though the overall state might, in fact, be assumed to be pure (i.e. at zero temperature). I present a simple derivation of the fluctuations formula as an instance of mixed states being reductions of higher level pure states, a representation that is affectionately known as ``Church of the Higher Hilbert Space". According to this view of mixed states, temperature is nothing but the amount of entanglement between the system and its environment.  
\end{abstract}

\maketitle

The purpose of this brief note is to put Jordan's derivation of Einstein's fluctuation formula into the modern quantum information setting. I will emphasise the role entanglement plays in this derivation even though the derivation took place ten years before Sch\"odinger introduced the concept of entanglement and called in the characteristic trait of quantum physics. Although Jordan's work is usually taken as the first step towards quantum field theory and second quantization (which it was), I will argue that it also hints at the representation of mixed states we call Church of the Higher Hilbert Space as well as the modern approch to thermodyanmics known as the ``eigenstate thermalisation hypothesis" (ETH). I will not pretend to be historically accurate. My account is simply physics-oriented. 

We will first present a derivation of the energy fluctuations in the black body in thermal equilibrium (this is how one would do it now, but Einstein didn't have access to most of this machinery). Black body is mathematically described as a collection of independent harmonic oscillators each representing a different mode of radiation. Because the modes are independent, in order to calculate the energy flucutations, it is enough to do so for just one mode (which we will take to be sinonimous with its frequency). In that case
\begin{equation}
\Delta_{\omega}^2 := (\hbar\omega)^2 (\langle (a^{\dagger}a)^2\rangle - \langle (a^{\dagger}a)\rangle^2)
\end{equation} 
which is straightforward to compute as it is basically the energy of the mode times the dispersion in the number of quanta present in that mode. One remembers that the average is performed with respect to the state
\begin{equation}
\rho_{\omega} = \frac{1}{e^{\beta H}-1}
\end{equation} 
where $H=\hbar\omega a^{\dagger}a$ (we omit the zero point energy). Then
\begin{equation}
\bar n_{\omega} :=  \langle (a^{\dagger}a)\rangle = \frac{1}{e^{\beta \hbar \omega}-1}
\end{equation} 
while (after a short calculation one finds that) 
\begin{equation}
\langle (a^{\dagger}a)^2\rangle = 2\bar n_{\omega}^2 + \bar n_{\omega} \;\; .
\end{equation} 
Therefore, the fluctuations can now be written in a simple way as
\begin{equation}
\Delta^2 \propto \bar n^2 + \bar n 
\end{equation} 
where we have omitted the subscript $\omega$ for simplicity. If we need the total fluctuations, we just need to perform the summation over all the relevant modes. Einstein, who derived this formula first (albeit without creation and annihilation operators or mixed states as this was well before the advent of the quantum mechanics proper \cite{Einstein}), noted that the first term in the despersion comes from the wave-like nature of the electromagnetic field comprising the black body while the second term comes from its particle-like tendencies. The fact that the fluctuations simply just add up led Einstein to (misleadingly) conclude that the waves and particles behave independently within the black body. This later became part of the so called wave-particle duality paradigm that featured strongly in Bohr's philisophy of quantum physics based on complementarity of the two descriptions. 

Shortly after Heisenberg wrote his groundbreaking paper on matrix mechanics, he, together with Born and Jordan, wrote a paper expanding on the basic rules of quantization \cite{Jordan}. The last chapter of that paper, known as Drei-M\"anner-Arbeit (the work of three men), is however exclusively due to Jordan \cite{Waerden}. In that chapter Jordan revisits Einstein's formula and derives it completely from the wave picture, albeit with a twist. The waves Jordan considers are collections of harmonic oscillators, but whose $x$ and $p$ components behave like matrices (just as Heisenberg had stipulated in his first paper). It is because the $x$'s and the $p$'s of these Harmonic oscillators do not commute that we obtain the second term of Einstein's formula. Otherwise, we would only have the first term which is a reflection of the wave-like nature of the classical electromagnetic field \cite{Born}.  In the modern language we would say that because the electric and the magnetic field components obey the standard commutation relations, the quantum electromagnetic energy fluctuations are different to their classical counterpart. This was completely clear to Jordan who emphasised that the result is a consequence of the quantum kinematics (i.e. the fact that $x$ and $p$ are operators and not just numbers) and not quantum dynamics (which was assumed to be the same as classical). 

Jordan considered this derivation his most important contribution to physics, however, it was viewed with suspicion by a number of his contemporaries (including Born and Heisenberg!) \cite{Duncan}. One problem was the divergence of the zero point energy (which Jordan introduced into quantum physics too) if summed up over all modes. This turns out to be irrelevant. Above we obtained the correct result by looking only at one frequency (and so there are no divergences), which is why we could even omit the zero point contribution without any loss (Jordan restricts himself to a narrow band of frequencies to the same effect). The second problem with Jordan's derivation, however, was that it was done with respect to number states of radiation (mixed states had not been invented for another two years). So, why did he claim to recover Einstein's expression? 

This is the part that I would like to focus on. Obviously Jordan knew that the energy fluctuations in an energy eigenstate are identically zero (by definition). But, if the total energy does not fluctuate, this means that the local energy (confined to a subvolume of the total volume) will have to fluctuate. In order to illustrate this phenomenon of local fluctuations when global ones are absent, let us look at a simple example where we have a single mode with $N$ quanta exactly. Assuming that it is divided into two spatial modes the state can be written as:
\begin{equation}
|\Psi\rangle = \sum_{n=0}^N c_n |n,N-n\rangle \;\; .
\label{2}
\end{equation}
It is clear that the total number of quanta in $|\Psi\rangle$ is always $N=n+(N-n)$ although each of the two spatial modes has an indeterminate number of quanta. Now we assume that the amplitudes have the form $\sqrt{e^{-an}/Z}$ where $Z=\sum e^{-an}$ and $a>0$ is some constant. In this case, the reduced states of each spatial mode approximate the thermal state (since their probabilities conform to the Boltzmann formula) with the density matrix given by
\begin{equation}
\rho = \sum_{n=0}^N |c_n|^2 |n\rangle\langle n|
\end{equation}
which, for large $N$, converges onto the thermal state. It is therefore clear that the number dispersion will be
\begin{equation}
\Delta^2 = (\sum_n |c_n|^2 n)^2 +  (\sum_n |c_n|^2 n) = \bar n^2 + \bar n 
\end{equation}
which is exactly Einstein's formula. Therefore, when two spatial modes (volumes) are entangled in an overall number state $N$, then each of them individually fluctuates in the number of partiels (and hence in energy) and the resulting fluctuation has a similar form to the thermal black body state (I say ``similar" because if the amplitudes are different, the coefficients in the dispersion pertaining to the linear and quadratic terms may change). This is not surprising since, when we trace out one of the modes, the other one ends up in a mixed state. The number fluctuations in this mixed state will always have both the linear as well as the quadratic term, i.e. particles and waves. If we want to attribute a (local) temperature to this mixed state, then this temperature will be a function of the amount of entanglement between the two spatial modes \cite{Vedral} (i.e. inversely proportional to the constant $a$ if we choose the amplitudes as above). 

I mention in passing that one can use the difference between the sum of the local dispersions of the two modes and the global dispersion as a measure of entanglement for pure bipartite states (see e.g. \cite{Amico} for quantification of entanglement). Only for product states, which are by definition not entangled, will this difference vanish. Otherwise it will always be positive and its value will depend on how entangled the modes are. 

The above argument contains the most important elements of Jordan's derivation, and, I will now present a more detailed calculation, but in the same simple spirit. It will allow us to reproduce the factor of $2$ exactly but for a very general (pure) state of bosons. The bunching property of bosons (in this case photons) will be seen to be responsible for that fact that $\langle (a^{\dagger})^2a^2\rangle = 2(\bar n)^2$ (and not just $=(\bar n)^2$), which was the crucial intermediate step leading to Jordan's version of Einstein's formula. 
So, let us (together with Jordan) assume that the overall state of the system is just a product of number states each pertaining to one frequency:
\begin{equation}
|\Psi\rangle \propto \prod_p a^{\dagger}(p)^{n(p)}|0\rangle  \;\; .
\end{equation}
In this state $n(p)$ is the number of photons in the mode $p$ (which we can think of either as the frequency, or as the momentum in which case it is related to the frequency via $pc=\hbar \omega$). This state is clearly the overall energy (or momentum) eigenstate and we would now like to caclulate the fluctuations of energy in a very small volume in space (we will actually compute this at a single point). The relevant quantity is  the bosonic pair correlation function defined as
\begin{equation}
\langle \Psi |\psi^{\dagger}(x)\psi^{\dagger}(y)\psi (x)\psi (y)|\Psi\rangle \;\; ,
\end{equation}
where $\psi (x) = \sum_p e^{ipx}a_p$ is the boson annihilation operator at the point $x$. In other words, we are computing the correlations between two spatial points $x$ and $y$ given that the overall state is $|\Psi\rangle$. This is just the ``infinitesimal" version of the calculation we did above with two spatial modes in eq. (\ref{2}). In the Fourier basis, we need to evaluate:
\begin{equation}
\langle \Psi |a^{\dagger}(p)a^{\dagger}(q)a(q') a(p')|\Psi\rangle \approx \sqrt{n(p)n(q)n(p')n(q')} (\delta (p-q)\delta (p'-q') + \delta (p-q')\delta (p'-q)) 
\end{equation}
which clearly contains the direct ($p\rightarrow q$ and $p'\rightarrow q'$) as well as the bosonic exchange ($p\rightarrow q'$ and $p'\rightarrow q$) contributions. The final result is the well known expression
\begin{equation}
\langle \Psi |\psi^{\dagger}(x)\psi^{\dagger}(y)\psi (x)\psi (y)|\Psi\rangle =  n^2 +|\int dp e^{ip(x-y)} n(p)|^2 \;\; ,
\end{equation}
where $n$ is the number denisty of photons (uniform in space for the state we are using). As $x \rightarrow y$, the above expression gives us the term that appears in the fluctuations (equal to $2n^2$) and is a clear indication of the bosonic bunching effect: the probability for bosons to occupy the same mode increases with the number of bosons. The dispersion in the state $|\Psi\rangle$ is now simple to calculate as we have all the relevant terms. It is straightforward to confirm that it has the Einsteinian form discussed before. We can always integrate this expression to obtain fluctuations in a small finite volume (rather than at a point $x$) but this won't change any of our conclusions. 

Just from the fact that he could recover Einstein's result, Jordan saw the crucial evidence for the intuition that the field has to be quantized by upgrading its relevant components into matrices. The field dynamics (basically the simple harmonic motion obeyed by each mode) did not need to be changed. Also, realising that when this is done one obtains the bosonic statistics was a great insight of Jordan's, leading many people to think of him as the father of the quantum field theory proper. Above we saw that the reason for the factor $2$ in the formula $\langle (a^{\dagger}a)^2\rangle = 2\bar n^2 +\bar n$ is actually what is now refered to as the bosonic bunching effect. Our initial calculation using a single mode in a thermal state concealed this fact since we need at least two frequencies to have the exchange interference effect. 

Magically, Jordan was not aware of the phenomenon of entanglement, or the notion of mixed states (he worked in the Heisenberg picture which completely marginalises states) or indeed the descriprions of fields in terms of creation and annihilation of particles. But the fact that he reproduced Einstein's earlier results means that Jordan inadvertendly discovered the notion of ``mixedness without mixedness" (or ``temperature without temperature" \cite{Vedral}). His derivation simply embodies the fact that a mixed partial state of a subsystem must arise from an overall pure state. This is why Jordan was also the forerunner of the Church of the Higher Hilbert Space perspective that is so ubiquitous in quantum information. 

Finally, I would like to mention that, because Jordan worked in the Heisenberg picture, his operators (representing $x$ and $p$ and thus the energy as well as its fluctuations) all evolved in time. He therefore had to perform time averages before averaging with respect to number states (which would not have added anything crucial to our discussion above). This is akin to the ETH approach to thermodyanmics where one relies on the observation that ``the individual energy eigenstates behave in many ways like a statistical ensemble" which is underpinned by the ergodic hypothesis (that time and ensemble averages coincide)  \cite{Deutsch}. Jordan's work can thus also be seen as the first contribution to the field of ETH.

\textit{Acknowledgments}: The author acknowledges funding from the National Research Foundation (Singapore), the Ministry of Education (Singapore) and Wolfson College, University of Oxford.

\end{document}